\documentstyle[mbf]{ptptex}
 \def\ind{\indent}
 \def\nn{\nonumber}

 \def\be{\begin{equation}}
 \def\ee{\end{equation}}
 \def\ben{\begin{enumerate}}
 \def\een{\end{enumerate}}
 \def\bl{\begin{flushleft}}
 \def\el{\end{flushleft}}
 \def\bt{\begin{tabular}}
 \def\et{\end{tabular}}
 
 \def\br{\begin{flushright}}
 \def\er{\end{flushright}}
 \def\bc{\begin{center}}
 \def\ec{\end{center}}
 \def\bea{\begin{eqnarray}}
 \def\eea{\end{eqnarray}}
 \def\bea*{\begin{eqnarray*}}
 \def\eea*{\end{eqnarray*}}
 \def\ba{\begin{array}}
 \def\ea{\end{array}}
 \def\bi{\begin{itemize}}
 \def\ei{\end{itemize}}
 
 \def\cD{{\mbox {${\cal D}$}}}

 \def\dslash{\partial{\raise 1pt\hbox{$\!\!\!/$}}}
 \def\cL{{\mbox {${\cal L}$}}}
 
 \def\cA{{\mbox {${\cal A}$}}}

 \def\cU{{\mbox {${\cal U}$}}}
 \def\bA{{\mbox {{\mbf A}}}}

 \def\bC{{\mbox {{\mbf C}}}}
 \def\bD{{\mbox {{\mbf D}}}}
 \def\bF{{\mbox {{\mbf F}}}}
 \def\bH{{\mbox {{\mbf H}}}}

 \def\cham{Chamseddine}
 \def\ind{\indent}
 \newfont{\bg}{cmr10 scaled\magstep4}
 \newcommand{\bigzerol}{\smash{\lower0.7ex\hbox{\bg 0}}}
 \newcommand{\bigzerou}{%
    \smash{\lower0.5ex\hbox{\bg 0}}}
 \newcommand{\bigzeroll}{\smash{\lower1.7ex\hbox{\bg 0}}}
 \newcommand{\bigzerouu}{%
    \smash{\lower1.7ex\hbox{\bg 0}}}
 \newcommand{\bigzerolll}{\smash{\lower2.5ex\hbox{\bg 0}}}
 \newcommand{\bigzerouuu}{%
    \smash{\lower2.5ex\hbox{\bg 0}}}
 \newcommand{\bigzeroL}{\smash{\lower0.01ex\hbox{\bg 0}}}

\def\be{\begin{equation}}
 \def\cD{{\mbox {${\cal D}$}}}

 \def\cL{{\mbox {${\cal L}$}}}
 \def\cA{{\mbox {${\cal A}$}}}
 \def\cU{{\mbox {${\cal U}$}}}

 \def\cU{{\mbox {${\cal U}$}}}
 \def\bA{{\mbox {{\mbf A}}}}

 \def\bC{{\mbox {{\mbf C}}}}
 \def\bF{{\mbox {{\mbf F}}}}
 \def\bH{{\mbox {{\mbf H}}}}

\makeatletter

          \@addtoreset{equation}{section}
       \makeatother
\title{%
Lagrangian Formulation of Connes' Gauge Theory
}
\vspace{5mm}
\author{%
  Hiromi {\sc Kase}, Katsusada {\sc Morita}$^{*}$ and Yoshitaka {\sc Okumura}$^{**}$
}
\inst{%
{\it  Department of Physics, 
  Daido Institute of Technology, Nagoya 457-0811, Japan\\
  $^{*}$Department of Physics, 
  Nagoya University, Nagoya 464-8602, Japan\\
  $^{**)}$ Department of Natural Sciences, 
  Chubu University, Kasugai, Aichi, 487-0027, Japan}
}

\recdate{%
\today
}
\vspace{10mm}

\abst{%
It is shown that
Connes' generalized gauge field
in non-commutative geometry is
derived
by simply requiring that
Dirac lagrangian be invariant under local transformations
of the unitary elements of
the algebra, which
define the gauge group.
The spontaneous breakdown of the
gauge
symmetry
is guaranteed provided
the chiral fermions exist
in more than one generations
as first observed by Connes-Lott.
It is also pointed out that
the
most general gauge invariant
lagrangian in the bosonic sector
has two more parameters
than in the original Connes-Lott scheme.
}
\begin{document}
\maketitle
\section{Introduction}
Connes' gauge theory\cite{1)} in non-commutative geometry (NCG)\cite{2),3)} 
based on Euclidean 4-space times
a 2-point space
employs the free Dirac operator with mass term
among others in the beginning
and unifies the gauge and Higgs fields
in an ingenious but highly mathematical
way.
A lot of works have been done
along this line of thought.
In view of so many papers
in this field
we may be excused to quote
only one recent article\cite{4)}.\\
\ind
On the other hand,
Sogami
proved\cite{5)} that it is possible to
describe the gauge and Higgs fields
in a unified way
using the generalized covariant derivative
defined for Dirac lagrangian with given
gauge-Yukawa interactions
for chiral fermions.
\\
\ind
One may then ask if
there exists
a lagrangian formulation of Connes' gauge theory,
which allows one 
to derive
Connes' generalized gauge field\cite{1),2)}
from a symmetry principle,
determining the type of interactions
for chiral fermions
with gauge and Higgs fields.
In the present paper we propose
such a lagrangian formulation.
\\
\ind
It is well known that
the free Dirac lagrangian
\be
\cL_D={\bar\psi}D\psi,
\psi=\left(
     \ba{l}
     \psi_L\\
     \psi_R\\
     \ea
     \right),\;
{\bar\psi}=({\bar\psi_L},{\bar\psi_R})
\label{eqn:1-1}
\ee
can be made gauge invariant if $\psi$ is non-chiral,
while it cannot be made chiral-gauge invariant
if $\psi$ is chiral;the reason is simply because the mass term
is chiral-non-invariant.
The free Dirac operator $D$ consists of
the free derivative operator and
the mass matrix\footnote{One may add
two arbitrary hermitian matrices to
the diagonal blocks
of the mass matrix.
For simplicity
we shall consider only the case indicated in the text.}
\be
D=D_0+i\gamma_5M,\;
D_0=\left(
    \ba{cc}
    i\gamma^\mu\partial_\mu&0\\
    0&i\gamma^\mu\partial_\mu\\
    \ea
    \right),\;
M=\left(
  \ba{cc}
  0&M_1\\
  M_1^{\dag}&0\\
  \ea
  \right),
\label{eqn:1-2}
\ee 
$\dag$ meaning the hermitian conjugation.
We take the Dirac matrices to satisfy $\{\gamma^\mu,\gamma^\nu\}=
2\eta^{\mu\nu},\;
\eta^{\mu\nu}={\rm diag}(+1,-1,-1,-1)$
with $\gamma_5=-i\gamma^0\gamma^1\gamma^2\gamma^3$.
The $\gamma_5$ matrix in front of the mass matrix is
only for later convenience.
The chiral spinors in Eq.(\ref{eqn:1-1}) are defined by
$\psi_{L,R}=2^{-1}(1\pm\gamma_5)\psi$.
\\
\ind
Usually
the gauge group is taken to be
any Lie group.
On the other hand,
in the Connes' approach\cite{1),2)}
the gauge group $G$ is 
regarded
as the unitary group of some (local) algebra \cA
\be
G=\cU(\cA)=\{g\in\cA;gg^{\dag}=g^{\dag}g=1\}.
\label{eqn:1-3}
\ee
The exceptional Lie groups
are then excluded from the gauge groups.
Moreover,
the well-known color gauge group $SU(3)$
cannot be obtained from a single algebra.
Instead
Connes' gauge theory can accommodate color symmetry
only if it is combined with flavor symmetry\cite{1),2),3)},
which points to a unification of color and flavor,
though the standard model gauge group is a product group.
Thus
considering an algebra
and its associated unitary group
seems to play a characteristic
role in gauge theory.
\\
\ind
The next section discusses a new style of symmetry principle
which, starting from Eq.(\ref{eqn:1-1}) for
given gauge group
(\ref{eqn:1-3}), determines the type
of interactions for chiral fermions
with gauge and Higgs fields.
Section 3 is devoted to discuss
the spontaneous breakdown of the symmetry
in the bosonic sector
which
is followed by Weinberg-Salam theory for leptons in section 4.
The last section gives a brief prescription how
to include quarks with color
and a short summary.
\section{Connes' generalized gauge field from symmetry principle}
\ind
We shall now demonstrate that
the Dirac lagrangian (\ref{eqn:1-1})
is easily made gauge invariant
irrespective of the chiral property of the fermions
if we employ the gauge group (\ref{eqn:1-3}).
Namely, we require that
the Dirac lagrangian 
be invariant
under the local transformation
$\psi\to a\psi$ for any invertible
element $a$
of the algebra \cA. (The set of
invertible elements
of the algebra forms a group.)
This is accomplished by replacing Eq.(\ref{eqn:1-1}) with
\be
\cL_{D}=\displaystyle{\sum_i}{\overline{(a_i\psi)}}D(a_i\psi),
\label{eqn:2-1}
\ee
where $a_i\in\cA$\footnote{Strictly speaking,
we should write $\rho(a)\psi$ for $a\psi$
and $\rho(a_i)$ for $a_i$ in Eq.(\ref{eqn:2-1}),
where the notation $\rho$ indicates the representation
of the algebra $\cA\;$ on the Hilbert space of the spinors.
For simplicity we omit the notation $\rho$ in what follows
unless necessary.}
and
the summation over the index $i$
is to be taken for later reason.
Since the transformation
$\psi\to a\psi$ 
is considered only for the invertible element $a$,
there exists $a^{-1}\in\cA$ such that the product
$a_ia^{-1}$ also belongs to the algebra
if $a_i\in\cA$.
Consequently,
the Dirac lagrangian
Eq.(\ref{eqn:2-1})
is invariant under the 
local transformation
$\psi\to a\psi$ for any invertible
element $a$
of the algebra $\cA$ provided that
all $a_i$ are transformed into $a_ia^{-1}$ at the same time.
We assume that any element of the algebra
does not change the chirality
so that it is regarded as  $2\times 2$ block-diagonal matrix
and
the same is also true for $g$.
\\
\ind
To further convert Eq.(\ref{eqn:2-1})
into gauge-invariant form,
we next require that the Lorentz scalar ${\bar\psi}\psi$ be also
invariant under the replacement $\psi\to a_i\psi$ together with the
summation over $i$. This leads to the condition
\be
\sum_ia_i^{\dag}a_i=1.
\label{eqn:2-2}
\ee
If the sum is reduced to a single term,
the involved element is necessarily
unitary.
Since we can choose any element from the algebra
to construct Eq.(\ref{eqn:2-1}),
this means that the algebra consists of unitary elements only,
which is impossible.
Hence we have to take the summation in Eq.(\ref{eqn:2-1}).
The condition (\ref{eqn:2-2}) which takes the same form as
in Ref. 6)
remains invariant
under $a_i\to a_ig^{\dag}$ if $g\in G=\cU(\cA)$.
This implies that the theory is now invariant
under the gauge transformation $\psi\to g\psi$
for $g\in G=\cU(\cA)$, 
which is simultaneously accompanied with
the transformation $a_i\to a_ig^{\dag}$,
where the representation content of the 
fermions is limited as we shall see in \S 4.\\
\ind 
Using Eq.(\ref{eqn:2-2}) and the Leibniz rule
we rewrite Eq.(\ref{eqn:2-1}) as
\be
\cL_{D}={\bar\psi}{\bD}\psi,
\label{eqn:2-3}
\ee
where the generalized covariant derivative is given by
\be
\bD=D+\bA,\;\;\bA=
\displaystyle{\sum_i}a_i^{\dag}[D,a_i].
\label{eqn:2-4}
\ee
The generalized
gauge field $\bA$
is essentially the same as
Connes' one\cite{1),2)}.
We henceforth
call it Connes' generalized gauge field.
We {\it derived} it from a new symmetry
principle
in the lagrangian formalism.
We have developed a similar idea but in the extended
differential formalism\cite{7)}.
The theory is automatically gauge invariant,
leading to the gauge transformation law
\be
\bA\to ^g{\!\!\bA}=
\displaystyle{\sum_i}ga_i^{\dag}[D,a_ig^{\dag}]
=g\bA g^{\dag}+g[D,g^{\dag}],
\label{eqn:2-5}
\ee
where use has been made of the condition (\ref{eqn:2-2}).
\\
\ind
At this point the theory is classified into two categories
depending on whether the fermion is non-chiral or chiral.
For non-chiral fermions the mass matrix commutes with
the gauge transformations, which means that
$M_1$ in Eq.(\ref{eqn:1-2}) is proportional to the unit matrix.
Hence it also commutes with any element of the algebra \cA,
reducing $\bA$ to the usual Yang-Mills gauge field
$A$,
\be
A=
\displaystyle{\sum_i}a_i^{\dag}[D_0,a_i]=i\gamma^\mu A_\mu.
\label{eqn:2-6}
\ee
If, on the other hand, 
the fermions are chiral,
$M$ no longer commutes with the element of the algebra \cA,
and the decomposition (\ref{eqn:1-2}) implies the corresponding decomposition
\be
\bA=A+i\gamma_5\Phi,\;\Phi=
\displaystyle{\sum_i}a_i^{\dag}[M,a_i].
\label{eqn:2-7}
\ee
That is, Connes' generalized gauge field $\bA$
unifies the ordinary gauge field $A$ with
the shifted Higgs field $\Phi$
if the fermions are chiral.
The usual Higgs field is defined by $H=\Phi+M$ which
transforms like $H\to gHg^{\dag}$.
We found that
this unification is already achieved at the lagrangian level
without recourse to 
NCG
based on Euclidean 4-space
times a 2-point space.
Thus
the lagrangian (\ref{eqn:2-3}) with Eq.(\ref{eqn:2-4}) determines
the type of interactions for chiral fermions
from our symmetry principle\footnote{The symmetry is 
broken down to the unbroken
subgroup consisting of those gauge transformations
which commute with the mass matrix.
The breakdown should be spontaneous
because
the 
transformation law of $\Phi$
is inhomogeneous,
$\Phi\to g\Phi g^{\dag}+g[M,g^{\dag}]$
unless $g$ commutes with the mass matrix.}.
\\
\ind
The $\gamma_5$ matrix in Eq.(\ref{eqn:2-3})
can easily be removed if we transform $\psi\to e^{i\pi\gamma_5/4}\psi$.
It then reads
\be
\cL_{\bD}={\bar\psi}\cD\psi,\;
\cD=D_0+A-H.
\ee
By writing $\cD=i\gamma^\mu\cD_\mu$
with use of the relation $\gamma^\mu\gamma_\mu=4$
Sogami
called\cite{5)} $\cD_\mu$ (with the so-called Sogami's term
added) the generalized covariant derivative.
He then proceeds to define the generalized field strength based on it
to obtain the correct bosonic lagrangian.
In our derivation which
determines
the underlying gauge-Yukawa interactions for chiral fermions,
Sogami's generalized
covariant derivative
is expressed in terms of the
auxiliary objects $a_i$.
Therefore, we cannot apply Sogami's method directly.
We shall instead follow the method developed in
Ref. 6)
where
the $\gamma_5$ matrix in Connes' generalized gauge field
plays a crucial role 
in determining
the bosonic lagrangian
below.
\section{Bosonic sector}
\ind
To complete the lagrangian formulation
we should consider the bosonic sector as well.
The field strength is defined by\cite{1),2),3),4),6)}
\be
\bF=d\bA+\bA^2,\;d\bA\equiv \displaystyle{\sum_i}[D,a_i^{\dag}][D,a_i],
\label{eqn:3-1}
\ee
which is gauge covariant, since
$d(^{g\!}\bA)=\sum_i[D,ga_i^{\dag}][D,a_ig^{\dag}]$ from Eq.(\ref{eqn:2-5})
and 
we have, again using the condition (\ref{eqn:2-2}),
\begin{eqnarray} 
^{g\!}\bF&=&d(^{g\!}\bA)+(^{g\!}\bA)^2\nn\\[2mm]
&=&\displaystyle{\sum_i}[D,ga_i^{\dag}][D,a_ig^{\dag}]-[D,g][D,g^{\dag}]-
[D,g]\bA g^{\dag}+g\bA[D,g^{\dag}]+g\bA^2g^{\dag}\nn\\[2mm]
&=&g(d\bA+\bA^2)g^{\dag}=g\bF g^{\dag}.
\label{eqn:3-2}
\end{eqnarray}
More elaborate proof without using the condition
(\ref{eqn:2-2}) was presented by Connes\cite{2)}.\\
\ind
It follows from Eqs.(\ref{eqn:2-7}) and (\ref{eqn:3-1}) that
we obtain 
\begin{eqnarray}
\bF&=&F-i\gamma_5[D_0+A,H]-Y,\nn\\[2mm]
Y&=&X+H^2-M^2-\displaystyle{\sum_i}a_i^{\dag}[M^2,a_i],\nn\\[2mm]
X&=&-\displaystyle{\sum_i}a_i^{\dag}\partial^2a_i
+\partial_\mu A^\mu+A_\mu A^\mu,
\label{eqn:3-3}
\end{eqnarray}
where $F=-(1/4)[\gamma^\mu,\gamma^\nu]F_{\mu\nu},\;
F_{\mu\nu}=\partial_\mu A_\nu-\partial_\nu A_\mu+[A_\mu,A_\nu]$.
The most general gauge-invariant lagrangian is given by
\be
\cL_B=-\displaystyle{{1\over 4}}{\rm Tr}\displaystyle{{1\over g^2}}\;{\tilde{\!\!\bF}}\bF.
\label{eqn:3-4}
\ee
Here the notation Tr means taking the
trace over the 2-dimensional chiral space,
Dirac matrices as well as internal symmetry
matrices and the gauge coupling constants become a diagonal
matrix $1/g^2$ commuting with the gauge transformations.
Also, an associated field strength $\;{\tilde{\!\!\bF}}\,$ 
is defined by
\be
{\tilde{\!\!\bF}}=\displaystyle{\sum_\alpha}h_\alpha^2\Gamma_\alpha\bF\Gamma^\alpha,
\label{eqn:3-5}
\ee
where $\alpha=S,V,A,T,P$
corresponding to $\Gamma_\alpha=1,\gamma_\mu,\gamma_5\gamma_\mu,\sigma_{\mu\nu}=(i/2)[
\gamma_\mu,\gamma_\nu],i\gamma_5$. 
We once proposed\cite{8)} similar
extention like
Eqs.(\ref{eqn:3-4}) and (\ref{eqn:3-5})
in Sogami's method.
Setting $\sum_\alpha h_\alpha^2
\Gamma_\alpha\sigma_{\mu\nu}\Gamma^\alpha
=\sigma_{\mu\nu}$ without loss of generality, 
we can see that $\;{\tilde{\!\!\bF}}\,$ has two more parameters than \bF
\be
{\tilde{\!\!\bF}}=F-\xi^2i\gamma_5[D_0+A,H]-\kappa^2Y
\label{eqn:3-6}
\ee
with $\xi^2$ and $\kappa^2$ being assumed to be positive.
(This is always possible since we have 5 parameters in
Eq.(\ref{eqn:3-5}).)
Needless to say,
$\xi^2=\kappa^2=1$ if we put $h_S^2=1, h_\alpha^2=0$ for $\alpha\not=S$.
Since the gauge transformation property
is not changed
for the sum (\ref{eqn:3-5}),
$\;{\tilde{\!\!\bF}}\,$ takes the same role as the field strength
$\bF\,$ in constructing the lagrangian. Hence we get
Eq.(\ref{eqn:3-4}).
\\
\ind
Substituting Eqs.(\ref{eqn:3-3}) and
(\ref{eqn:3-6})
into Eq.(\ref{eqn:3-4}) and
using the property of the trace of Dirac matrices,
we find that
\be
\cL_B=\cL_{YM}+\xi^2{\rm tr}\displaystyle{{1\over g^2}}(D^\mu H)(D_\mu H)
-\kappa^2{\rm tr}\displaystyle{{1\over g^2}}Y^2,
\label{eqn:3-7}
\ee
where $\cL_{YM}$ is the Yang-Mills lagrangian,
$D_\mu H=[\partial_\mu+A_\mu,H]$ and
the notation tr now means taking the
trace over the 2-dimensional chiral space
and internal symmetry
matrices.
Equation of motion for the auxiliary field in $Y$
reads $Y=0$ and the last term in Eq.(\ref{eqn:3-7})
vanishes identically.
This result is well known\cite{6)},
producing no Higgs potential upon 
elimination of the auxiliary field.
The authors in Ref.6)
evaded this unpleasant situation by
assuming that
fermions exist in generation
with nontrivial generation mixing.\\
\ind
This observation should be linked with
the fact that
the representation content of the fermions
is limited by
an underlying algebra representation.
\section{Weinberg-Salam theory in the leptonic sector}
\ind
As an illustration let us consider Weinberg-Salam theory in the leptonic 
sector. 
We shall see that
the flavor algebra to be considered below
allows only doublets and singlets
and, moreover,
quarks with fractional charges cannot be included unless color
is taken into account.
\\\ind
The flavor algebra for leptons
is taken to be
$\cA=C^\infty(M_4)\otimes(\bH\oplus\bC)$
,
where $C^\infty(M_4)$ denotes the set of
infinitely many differentiable
functions over Minkowski space-time $M_4$,
$\bH$ the real quaternion 
and $\bC$ the complex field.
The unitary group is $G=\cU(C^\infty(M_4)\otimes(\bH+\bC))
={\rm Map}(M_4,SU(2)\times U(1))$.
Choosing the mass-eigenstates basis
\be
\psi_L=\left(
               \ba{c}
               \nu_e\\
               U_le\\
               \ea
               \right)_L,\; 
               \psi_{eR}=\left(
               \ba{c}
               \nu_{eR}\\
               e_R\\
               \ea
               \right),
\label{eqn:4-1}
\ee
where generation indices are omitted
and $U_l$ is the leptonic Kobayashi-Maskawa matrix
(we assume massive neutrinos),
we consider the representation\cite{7)}
\be
\rho(a,b)=
         \left(
         \ba{cc}
         a&0\\
         0&B\\
         \ea
         \right)\otimes 1_{N_g},\;
         B=\left(
           \ba{cc}
           b&0\\
           0&b^{*}\\
           \ea
           \right),
\label{eqn:4-2}
\ee
where $a=a(x)\in C^\infty(M_4)\otimes\bH$,
$b=b(x)\in C^\infty(M_4)\otimes\bC$ with
$b^{*}$ being the complex conjugate to $b$,
so that left-handed and right-handed fermions in generation
belong to the doublet and singlet
of the gauge group, respectively.
This statement is obtained
by letting
the element $(a,b)$ belong to the unitary group of the algebra.
Remember that
the real quaternion has only one irreducible
representation of dimensions 2.
The matrix $B$ cannot be equal to $a$ or $a^{*}$
because, in that case,
both $\psi_{L,R}$ are doublets with no $U(1)$ charge,
leading to vector-like theory.
Moreover,
it must take the form
of Eq.(\ref{eqn:4-2}) or $b\leftrightarrow b^{*}$,
because the $2\times 2$ matrix-valued
Higgs field $h$ of
Eq.(\ref{eqn:4-4}) below
receives the gauge transformation,
$h\to ahB^{\dag}$ for unitary matrices $(a,B)$.
This automatically determines
the correct
hypercharge assignment
of Higgs field.
The representation (\ref{eqn:4-2}), however, gives rise to
wrong hypercharge assignment of
leptons.
It turns out that
to remedy this point
without changing
the correct
hypercharge
of Higgs
will require only a minor change in the
theory.
The matrix $1_{N_g}$ is the $N_g$-dimensional
unit matrix here in generation space.\\
\ind
To show how Higgs potential naturally appears
in the present formalism,
we choose the mass matrix
\be
M=\left(
  \ba{cc}
  0&M_1\\
  M_1^{\dag}&0\\
  \ea
  \right),\;\;M_1=\left(
                  \ba{cc}
                  m_1&0\\
                  0&m_2\\
                  \ea
                  \right),\;m_{1,2}:\;N_g\times N_g
\label{eqn:4-3}
\ee
to obtain Higgs field as\footnote{We have used the condition
(\ref{eqn:2-2}) in the present case
to show that $\varphi^{\dag}$
in Eq.(\ref{eqn:4-4})
is the hermitian conjugate to $\varphi$.} 
\begin{eqnarray}
\Phi&=&\left(
     \ba{cc}
     0&\varphi M_1\\
     M_1^{\dag}\varphi^{\dag}&0\\
    \ea
    \right),\;\;\varphi=\left(
                  \ba{cc}
                  \varphi_0^{*}&\varphi_+\\
                  -\varphi_-&\varphi_0\\
                  \ea
                  \right),\nn\\[2mm]
H&=&\Phi+M=\left(
         \ba{cc}
         0&hM_1\\
         M_1^{\dag}h^{\dag}&0\\
         \ea
         \right),\;
h=\varphi+1_2=\left(
             \ba{cc}
             \phi_0^{*}&\phi_+\\
             -\phi_-&\phi_0\\
             \ea
             \right).
\label{eqn:4-4}
\end{eqnarray}
\ind
It can be shown that Eq.(\ref{eqn:2-3}) with
(\ref{eqn:2-4}) leads to, after transforming
$\psi\to e^{i\pi\gamma_5/4}\psi$,
\begin{eqnarray}
\cL_D&=&{\bar\psi}_Li\gamma^\mu(\partial_\mu-\displaystyle{{ig_2\over 2}}
A_\mu^{(2)})\psi_L+{\bar\psi}_Ri\gamma^\mu
(\partial_\mu-\displaystyle{{ig_1\over 2}}
\left(
\ba{cc}
1&0\\
0&-1\\
\ea
\right)
A_\mu^{(1)})\psi_R\nn\\[2mm]
&&\;\;-{\bar\psi}_L(m_1{\tilde\phi},m_2\phi)\psi_R
-{\bar\psi}_R\left(
             \ba{cl}
             {\tilde\phi}^{\dag}m_1^{\dag}\\
             \phi^{\dag}m_2\\
             \ea
             \right)
             \psi_L,\;{\tilde \phi}=\left(
                                    \ba{cc}
                                    0&1\\
                                    -1&0\\
                                    \ea
                                    \right)\phi^{*},
\label{eqn:4-5}
\end{eqnarray}
where
\be
\phi=\left(
     \ba{cl}
     \phi_+\\
     \phi_0\\
     \ea
     \right),\;\;\langle\phi\rangle=\left(
                                    \ba{cl}
                                    0\\
                                    1\\
                                    \ea
                                    \right)
\label{eqn:4-6}
\ee
is the normalized Higgs field,
and we set
\begin{eqnarray}
A&=&\displaystyle{\sum_i}\rho(a_i^{\dag},b_i^{\dag})
[D,\rho(a_i,b_i)]\nn\\[2mm]
&&
=i\gamma^\mu
 \left(
 \ba{cccc}
 -\displaystyle{{ig_2\over 2}}A_\mu^{(2)}& & &\bigzerou\\
 \bigzerol& & &-\displaystyle{{ig_1\over 2}}\left(
                 \ba{cc}
                                1&0\\
                                0&-1\\
                                \ea
                                \right)A_\mu^{(1)}\\
 \ea
 \right)\otimes 1_{N_g}.
\label{eqn:4-7}
\end{eqnarray}
Here, $A_\mu^{(2)}$ is an $SU(2)$ gauge field,
while  $A_\mu^{(1)}$ is a $U(1)$ gauge field. Both are hermitian.
Also, $g_2$ and $g_1$ are the respective gauge coupling constants.
\\
\ind
Let us now evaluate the bosonic lagrangian (\ref{eqn:3-7}).
Writing
\be
\displaystyle{{1\over g^2}}
=\displaystyle{{4\over N_g}}
        \left(
        \ba{cc}
        g_2^{-2}1_2&0\\
        0&g_1^{-2}1_2\\
        \ea
        \right)\otimes 1_{N_g},
\label{eqn:4-8}
\ee
we have, upon eliminating 
the auxiliary fields in tr$Y^2$
by the equation of motion
and rescaling the Higgs field
$\phi\to{1\over\xi\sqrt{L}}\phi={\sqrt{2}\over v}\phi$,
which renders the Yukawa term in Eq. (\ref{eqn:4-5}) multiplied by
${\sqrt{2}\over v}$,
\be
\cL_B=-\displaystyle{{1\over 8}}{\rm tr}F_{\mu\nu}^{(2)}F^{(2)\mu\nu}
-\displaystyle{{1\over 4}}F_{\mu\nu}^{(1)}F^{(1)\mu\nu}
+(D^\mu\phi)^{\dag}(D_\mu\phi)
-\displaystyle{{\lambda\over 4}}
(\phi^{\dag}\phi-\displaystyle{{v^2\over 2}})^2,
\label{eqn:4-9}
\ee
where
$D_\mu\phi=(\displaystyle{\partial_\mu}
-\displaystyle{{ig_2\over 2}}A_\mu^{(2)}
-\displaystyle{{ig_1\over 2}}A_\mu^{(1)})\phi$, and we define
\begin{eqnarray}
\lambda&=&\displaystyle{{4\kappa^2K\over \xi^4L^2}},\;\;
v^2=2\xi^2L,\;
L=\displaystyle{{2\over N_g}}(\displaystyle{{1\over g_2^2}}+
\displaystyle{{1\over g_1^2}}){\rm tr}_g(M_1^{\dag}M_1),\nn\\[3mm]
&&
K=\displaystyle{{2\over N_gg_1^2}}({\rm tr}_g(M_1^{\dag}M_1)^2
-\displaystyle{{1\over N_g}}({\rm tr}_gM_1^{\dag}M_1)^2)
\label{eqn:4-10}
\end{eqnarray}
with tr$_g$ denoting the trace in the generation space.
\\
\ind
It follows that
the symmetry breaking
occurs only if
$N_g>1$ as in Ref.6).
This result
was 
first observed by Connes-Lott\cite{1)}
in connection with their reconstruction of the
standard model in NCG.
It should be recalled that
our reconstruction has two more parameters than in
Connes-Lott scheme\cite{1)},
whence
our lagrangian (\ref{eqn:4-9})
is renormalizable with no constraint
among the bare parameters.
\\
\ind
As alluded to above
the hypercharge assignment of leptons
according to the present representation
(\ref{eqn:4-2})
is not phenomenologically correct:
$Y=0$ for $\psi_L$,
$Y=+1$ for $\nu_R$,
$Y=-1$ for $e_R$,
leading to
charge +1/2 for $\nu$ and $-1/2$ for $e$,
but the hypercharge of Higgs is correctly given, $Y=1$.
To obtain the correct hypercharge assignment
of leptons
we simply double the chiral spinors 
by including the charge conjugate spinor $\psi^c$ as well
\be
\Psi=\left(
     \ba{c}
     \psi\\
     \psi^c\\
     \ea
     \right),\;
\psi=\left(
     \ba{c}
     \psi_L\\
     \psi_R\\
     \ea
     \right),\;
     \psi_L=\left(
            \ba{c}
            \nu\\
            e\\
            \ea
            \right)_L,\;
     \psi_R=\left(
            \ba{c}
            \nu_R\\
            e_R\\
            \ea
            \right),
\label{eqn:4-11}
\ee
and consider the representation
\be
\rho(a,b)=\left(
           \ba{ccc}
          a&     &\bigzerouuu  \\
          \bigzerolll  &B \\
           &     &b1_4\\
           \ea
           \right)
          \otimes 1_{N_g}\equiv \rho(a,B;b1_2,b1_2).
\label{eqn:4-12}
\ee
This time the gauge transformation on the spinor $\Psi$
is not given by simply making the element
$(a,b)$ in Eq. (\ref{eqn:4-12})
unitary
because
the
gauge transformation on $\psi$
determines
that on $\psi^c$
by
the charge conjugation.
In fact, it is given by
the product
$\rho(a,B;b1_2,b1_2)\rho(b^{*}1_2,b^{*}1_2;a^{*},B^{*})$
for unitary element $(a,b)$ of the
flavor algebra.
Although
this is not an algebra representation,
Connes' generalized gauge field
allows\cite{3)} to construct
gauge-invariant theory also in this case
using
the real structure.
In what follows we shall simplify Connes'
presentation\cite{3)}
in the lagrangian formalism.\\
\ind
Since the additional 4-dimensional submatrix in Eq. (\ref{eqn:4-12})
is proportional to the unit matrix,
there is no change in the Higgs sector\footnote{Conversely,
there must be no change in the Higgs sector
because the hypercharge is already
correct, $Y=1$ for Higgs.
This necessarily implies
that the additional 4-dimensional
matrix is proportional to the unit
matrix $b1_4$ or $b^{*}1_4$.
What determines $b$ or $b^{*}$
is the requirement that
$\nu_{eR}$ be $U(1)$-neutral,
leading to $b$ for $B$
of Eq. (\ref{eqn:4-2}).}.
The fact that only the representation like
$B$ or $b1_4$
occurs implies the quantization
of $U(1)$ charges to be $\pm 1$
in this model.
Hence fractional charges can
arise only from another
stuff.
In this case we have to replace $A$ in Eq.(\ref{eqn:4-7})
with 
\be
\left(
\ba{cc}
A&0\\
0&A^c\\
\ea
\right),\;
A^c=-\displaystyle{{ig_1\over 2}}
    A_\mu^{(1)}1_4\otimes 1_{N_g}.
%    \left(
%          \ba{cccc}
%          1& &\bigzerouu & \\
%          &1& & \\
%          \bigzeroll& &1& \\
%          & & &1\\ 
%          \ea
%          \right)\otimes 1_{N_g}.
\label{eqn:4-13}
\ee
Using the elementary formula 
${\bar{\psi^c}}A^c\psi^c={\bar\psi}A^{c*}\psi$,
where $A^{c*}$ is complex conjugate to $A^c$,
we obtain the following
Dirac lagrangian instead of Eq.(\ref{eqn:4-4})
\begin{eqnarray}
\cL_D&=&{\bar\psi}_Li\gamma^\mu(\partial_\mu-\displaystyle{{ig_2\over 2}}
A_\mu^{(2)}-
\displaystyle{{ig_1\over 2}}
\left(
\ba{cc}
-1&0\\
0&-1\\
\ea
\right)
A_\mu^{(1)})\psi_L\nn\\[2mm]
&&\;\;+{\bar\psi}_Ri\gamma^\mu(\partial_\mu-\displaystyle{{ig_1\over 2}}
\left(
\ba{cc}
0&0\\
0&-2\\
\ea
\right)
A_\mu^{(1)})\psi_R
+{\rm Yukawa}{\mbox{-}}{\rm terms}.
\label{eqn:4-14}
\end{eqnarray}
We thus obtain the correct hypercharge       
assignment(delete $\nu_{eR}$
from Eq.(\ref{eqn:4-14}) if necessary).
The bosonic sector is untouched in this process
because the hypercharge assignment of Higgs doublet
is the same as before
and no additional gauge bosons appear.
\section{Summary}
\ind
The Weinberg-Salam theory in the previous section
is applicable only to leptons because
$U(1)$ charge is quantized to be $\pm 1$.
Quarks have fractional $U(1)$ charges
and can only be incorporated into the scheme
by taking account of color.\\
\ind
The color-flavor algebra\cite{3),4)} is
$\cA=C^\infty(M_4)\otimes(\bH\oplus\bC\oplus M_3(\bC))$,
where 
$M_3(\bC)$ is the set of $3\times3$ complex matrices,
which is represented on the doubled spinor (\ref{eqn:4-11})
with 
\be
\psi_L=\left(
      \ba{c}
      q_L\\
      l_L\\
      \ea
      \right),\;\;
\psi_R=\left(
      \ba{c}
      u_R\\
      d_R\\
      \nu_R\\
      e_R\\
      \ea
      \right),\;\;
      q_L=\left(
          \ba{c}
          u\\
          U_qd\\
          \ea
          \right),\;\;
          l_L=\left(
          \ba{c}
          \nu\\
          U_le\\
          \ea
          \right)
\label{eqn:5-1}
\ee
($U_q$ being the Kobayashi-Maskawa matrix) as\cite{3),4)}
\begin{eqnarray}
\rho(a,b,c)&=&\left(
            \ba{cc}
            \rho_w(a,b)&0\\
            0&\rho_s(b,c)\\
            \ea
            \right)\otimes 1_{N_g},\nn\\[2mm]
&&\rho_w(a,b)=\left(
                \ba{cc}
                a1_4&0\\
                0&B1_4\\
                \ea
                \right),\nn\\[2mm]
&&\rho_s(b,c)=\left(
                \ba{cccc}
                1_2\otimes c^{*}& & &\bigzerol\\
                &b1_2 &&  \\
                & & 1_2\otimes c^{*}&\\
                \bigzeroL& & &b1_2\\
                \ea
                \right),
\label{eqn:5-2}
\end{eqnarray}
where $c=c(x)\in C^\infty(M_4)\otimes M_3(\bC)$
and $(a,b)$ is the same as before.
The hypercharges of the leptons remain unchanged,
while those of quarks receive also from the phase of $3\times 3$
complex matrix $c$ defined at each space-time point.
One can assume\cite{7)}
det$\rho(a,b,c)=$det$\rho_s(b,c)=1$
for unitary element $(a,b,c)$
of the color-flavor algebra
in accordance with
Connes' unimodularity condition\cite{3),4)},
which 
implies
that
the hypercharge coming from $c$ is 1/3.
Looking at the representations (\ref{eqn:4-13})
as well as
(\ref{eqn:5-2})
the hypercharges of quarks are given by the sum
of those of the corresponding leptons with $1+1/3=4/3$.
This gives the correct hypercharge assignment
of quarks.
For instance,
$Y=-1+4/3=1/3$ for $q_L$,
$Y=0+4/3=4/3$ for $u_R$
and
$Y=-2+4/3=-2/3$ for $d_R$,
thereby assuring the anomaly cancellation
in each generation.\\
\ind
We shall no longer dwell upon the details
of the standard model in the present scheme
since it has repeatedly 
been discussed in the literature.
See, for instance, Refs. 4) and 7).
We have to concede, however, 
that
we have no quantitative idea yet
about what the auxiliary objects $a_i$
represent physically.\\
\ind
The present paper
has introduced an elementary
method of reformulating Connes' gauge theory
from the new style of symmetry principle
in the lagrangian formalism.
Using the present method,
one can apply
Connes' gauge theory to the
particle models
without resort to NCG.
This would greatly facilitate the model buildings
along this line of thought.
As an application
we shall
discuss
gravity
in the following paper.
%\vspace{5mm}
\section*{Acknowledgements}
One (K.M.) of the authors is grateful to
Professor S. Kitakado
for useful discussions and
continuous encouragement.
%\vspace{5mm}
%\newpage

\end{document}